# Clouds of short-circuited thermionic nanobatteries and promising prospects for development of nanobattery-based aerosol fusion reactors. The preliminary report.

Oleg Meshcheryakov

*Wing Ltd Company, 65000, 33 French blvd., Odessa, Ukraine, e-mail: wing99@mail.ru*

**Abstract**

The physical mechanisms of periodic separation and relaxation of electric charges within aerosol particles possessing the properties the short-circuited batteries can be extremely diverse. With use of appropriate materials and dispersing methods, the electrochemical, thermoelectric, thermionic, pyroelectric, photoelectric, photo electronic emission, or even radionuclide-based emission micro and nano-batteries can be synthesized and be dispersed in the air as clouds self-assembed of the short-circuited aerosol batteries due to the inter-particle electromagnetic dipole-dipole attraction. Intense thermionic emission from ionized hot spots migrating on the relatively cold surface of charged explosive particles, can convert these particles into short-circuited thermionic batteries, turning an aerosol cloud consisting of such unipolar charged, gradually decomposing explosive particles into ball lightning. The slow exothermic decomposition of the highly sensitive explosive aerosol particles, catalyzed by excess ions on their surface, and also ion-catalyzed reactions of slow water vapor induced oxidation of charged combustible aerosol particles underlie two main classes of natural ball lightning. At the same time, the artificially generated clouds consisting of such unipolar charged aerosol nanobatteries, probably, can have some useful applications, not only military ones. In particular, it seems that high-performance pyroelectric fusion reactors could be created on the basis of such ball-shaped aerosol clouds self-assembled of pyroelectric nanocrystals - short-circuited pyroelectric nanobatteries. Being condensed from laser-, microwave- or arc- produced vapor and being freely dispersed in the atmosphere containing hydrogen, deuterium or tritium under a millitorr pressure, the unipolar charged pyroelectric aerosol nanocrystals of lithium niobate or lithium tantalate could be rapidly radiatively cooled, thereby be strongly polarized. Thus, they could form the unipolar charged, centrally polarized ball cloud, in which repetitive collective pyroelectric generation of a strong, centrally symmetric electric field contributes to repetitive events of intense gas ionization , radial ion accelerating and highly exothermic fusion of the mutually colliding light nucleus of deuterium and tritium or hydrogen and lithium. The continuously repetitive cycles of the consecutive recooling, fusion-induced reheating, and repolarizing the unipolar charged pyroelectric aerosol nanocrystals in the ball-shaped cloud could be the basis of energy generation in such aerosol fusion reactors.

The model of ball lightning [1, 2] as an aerosol cloud consisting of unipolar charged particles, each of which possesses the properties of a tiny short-circuited electric generator (a short-circuited aerosol battery), and some possible practical applications of this model are developed here. The model easily explains an apparent intense mutual attraction between unipolar charged elements of a ball lightning substance as a mutual electromagnetic dipole-dipole attraction that always takes place between periodically or continuously short-circuited individual electric batteries. Also, the model easily explains violent electromagnetic phenomena, often accompanying ball lightning, in terms of collective electromagnetic effects occurring in the small unipolar charged cloud consisting of trillions of the tiny electric aerosol generators.

The physical mechanisms of periodic separation and relaxation of electric charges within the aerosol particles possessing the properties the short-circuited batteries can be extremely diverse. With use of appropriate materials and dispersing methods, the electrochemical, thermoelectric, thermionic, pyroelectric, photoelectric, photoelectronic emission, or even radionuclide-based emission micro and nano-batteries can be synthesized and be dispersed in the air as clouds self-assembed of the short-circuited aerosol batteries due to the inter-particle electromagnetic dipole-dipole attraction. The internal and external energy sources can cause the periodic separation and relaxation of electric charges within aerosol particles or on their surface. In particular, the internal or external energy sources can contribute to periodic or continuous thermionic emission from hot spots that migrate on the surface of such particles, thereby transforming these particles into thermionic aerosol batteries.

It is natural that aerosol particles, emitting electrons from their surface, always acquire a small equilibrium positive charge. However, the aerosol particles capable of emitting electrons can also be pre-charged with a surplus positive charge, e.g., during forming the particles in the zone of corona discharge. And so, a strong additional Coulomb repulsion will arise between such unipolar charged particles emitting electrons. If the aerosol particles that emit electrons are very hot and/ or if they are releasing gases from their surface, thermophoresis and diffusiophoresis will also contribute to mutual repulsion between such particles at short distances. The dynamic competition between the total forces of interparticle repulsion and forces of interparticle electromagnetic dipole-dipole attraction eventually will form a ball-shaped cloud of these unipolar charged, short-circuited aerosol batteries [1]. Despite the above diversity of possible types of the short-circuited aerosol batteries, here we would only like to focus on the mechanisms of the collective exothermic electrogenerating in the ball-shaped luminous aerosol clouds containing the unipolar charged particles, each of which possesses the properties of the short-circuited thermionic battery, emitting heat, photons and thermoelectrons from hot spots randomly migrating on its surface.

Intense thermionic emission from ionized hot spots migrating on the relatively cold surface of charged explosive particles, can convert these particles into short-circuited thermionic batteries, turning an aerosol cloud consisting of such unipolar charged, gradually decomposing explosive particles into ball lightning. The slow exothermic decomposition of the highly sensitive explosive aerosol particles, catalyzed by excess ions on their surface, and also ion-catalyzed reactions of slow water vapor induced oxidation of charged combustible aerosol particles underlie two main classes of natural ball lightning. At the same time, the artificially generated clouds consisting of such unipolar charged aerosol nanobatteries, probably, can have some useful applications, not only military ones. In particular, it seems that high-performance pyroelectric fusion reactors could be created on the basis of such ball-shaped aerosol clouds self-assembled of pyroelectric nanocrystals - short-circuited pyroelectric nanobatteries. Being condensed from laser-, microwave- or arc- produced vapor and being freely dispersed in the atmosphere containing hydrogen, deuterium or tritium under a millitorr pressure, the unipolar charged pyroelectric aerosol nanocrystals of lithium niobate or lithium tantalate could be rapidly radiatively cooled, thereby be strongly polarized. Thus, they could form the unipolar charged, centrally polarized ball cloud, in which repetitive collective pyroelectric generation of a strong, centrally symmetric electric field contributes to repetitive events of intense gas ionization , radial ion accelerating and highly exothermic fusion of the mutually colliding light nucleus of deuterium and tritium or hydrogen and lithium. The continuously repetitive cycles of the consecutive recooling, fusion-induced reheating, and repolarizing the unipolar charged pyroelectric aerosol nanocrystals in the ball cloud could be the basis of energy generation in such aerosol fusion reactors.

The extremely developed surface of the continuously repolarizing pyroelectric aerosol nanocrystals could contribute to effective near-surface ionization of the surrounding gas containing hydrogen isotopes. Also, the highly developed surface of the pyroelectric nanocrystals could contribute to ultrafast processes of their repetitive radiative cooling and fusion-induced heating in the repolarizing ball-shaped cloud. It is clear that if the average intra-cloud strength of the radial electric field in the ball-shaped cloud with a diameter of about 20 centimeters, consisting of the unipolar charged pyroelectric / thermionic aerosol nanobatteries, is of ~ 10 – 20 kV/cm, the hydrogen isotopes can be readily ionized in the immediate vicinity of the repolarizing surface of the pyroelectric aerosol crystals, and correspondingly the nuclei of the hydrogen isotopes can be readily accelerated in such pulsating radial intra-cloud electric fields up to critical energies of ~ 200 – 400 keV, with the possible total energy of their mutual colliding of ~ 400 – 800 keV.

Several specific recipes of experimental generation of ball lightning as the self-assembled luminous aerosol cloud consisting of the unipolar charged, short-circuited thermionic batteries formed, for example, of exothermically decomposing droplets of pure hydrogen peroxide, along with some illustrations of our experiments (see, e.g., Figures 1-11), will be shown in the next section of this report.

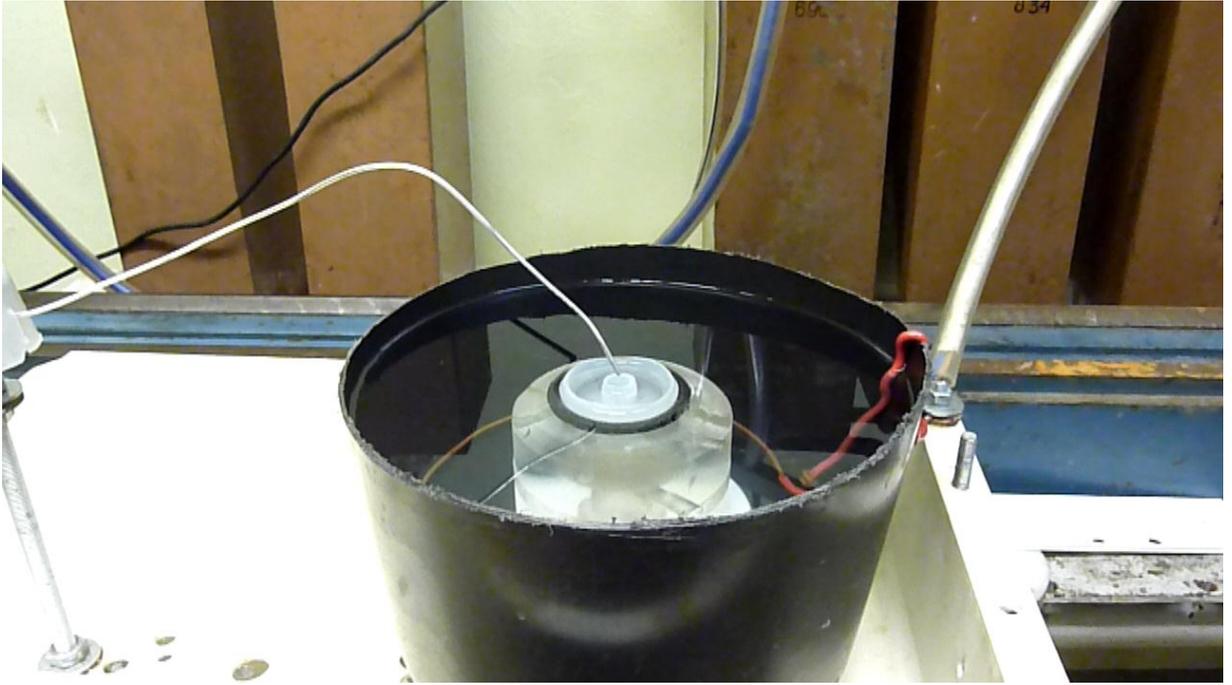

Fig.1

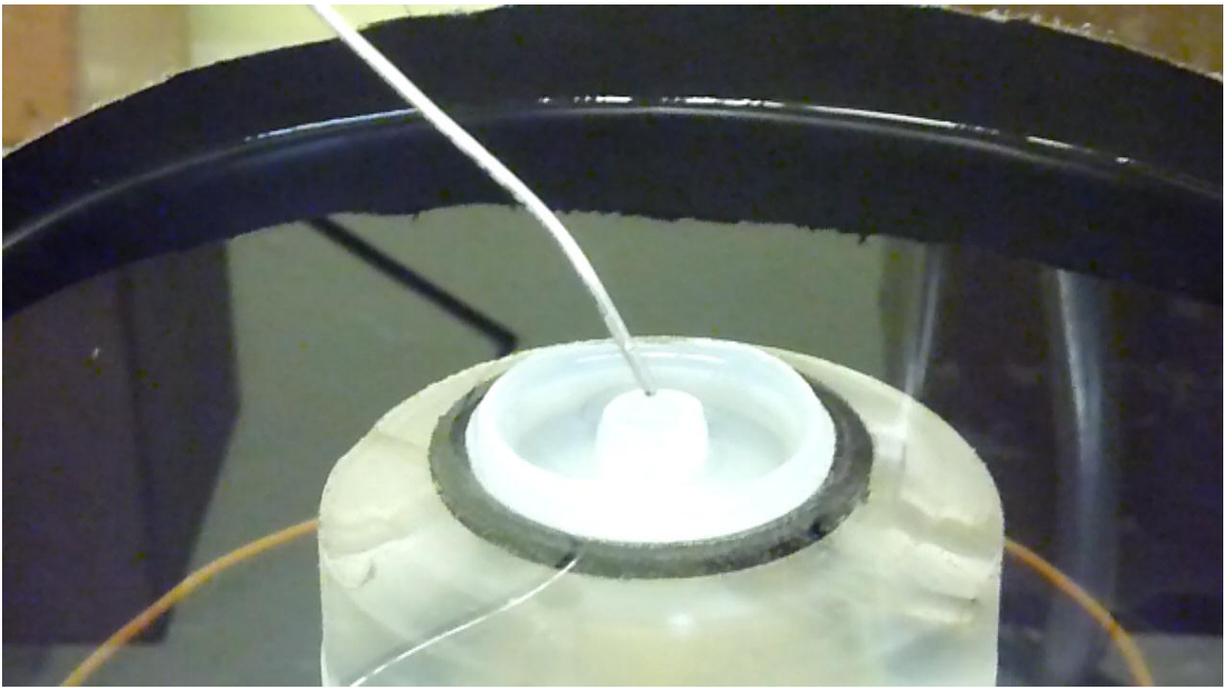

Fig.2

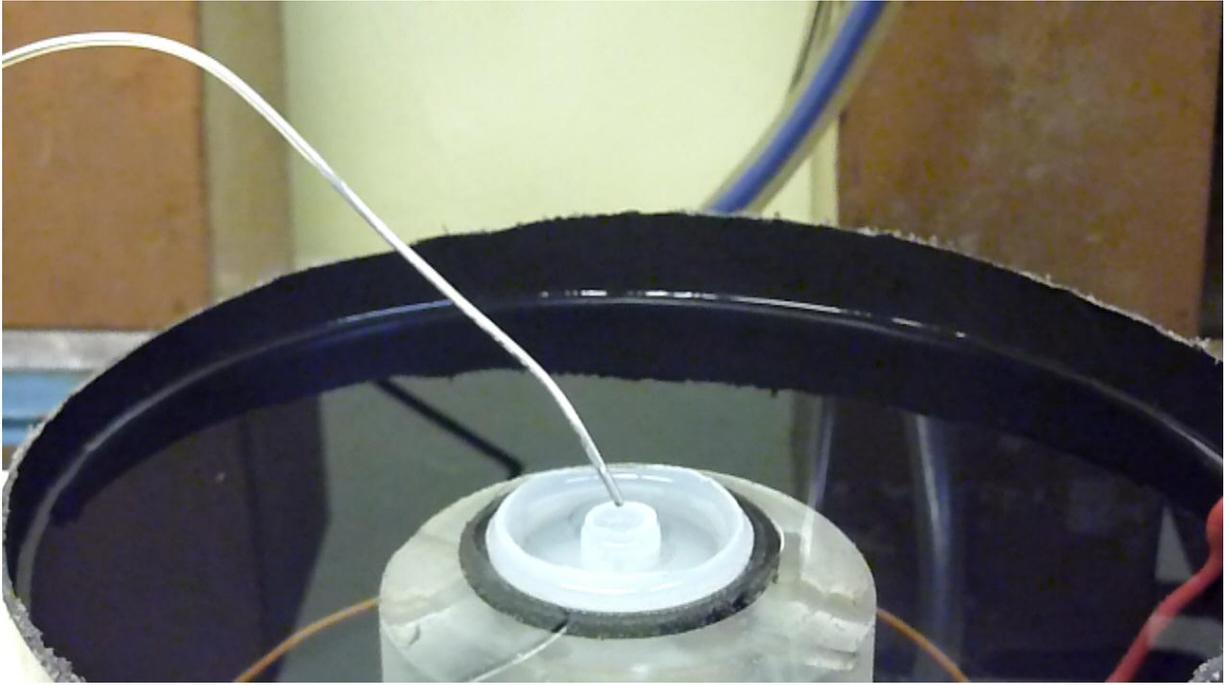

Fig.3

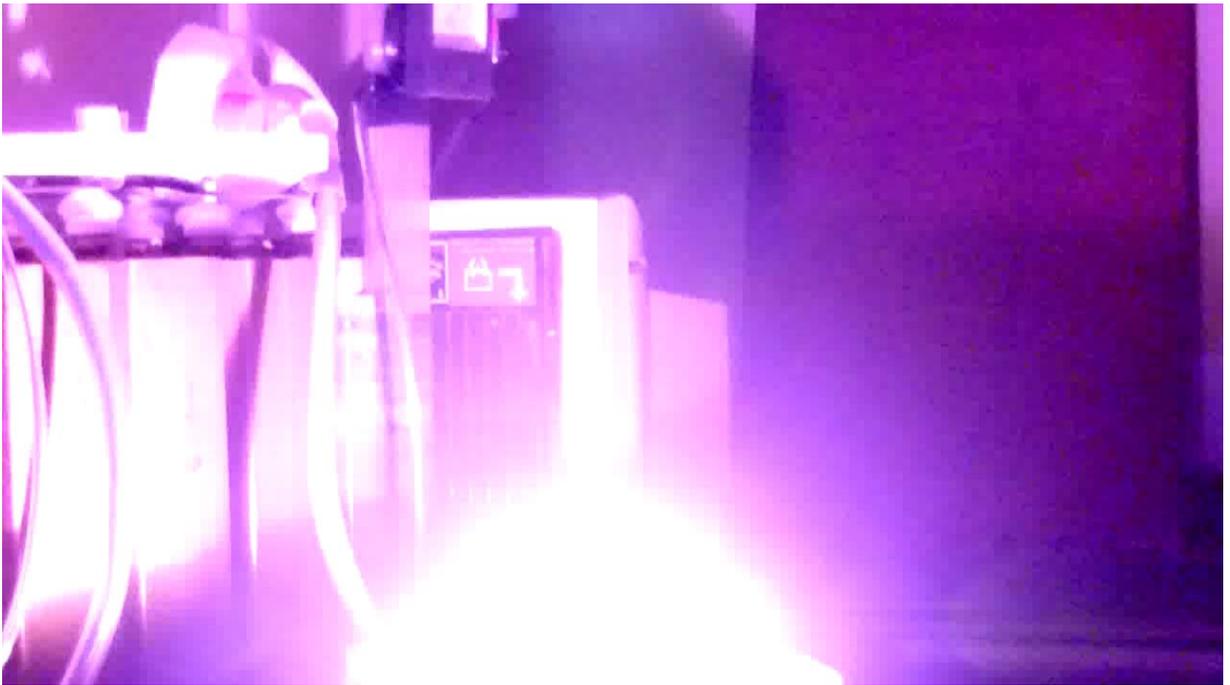

Fig.4

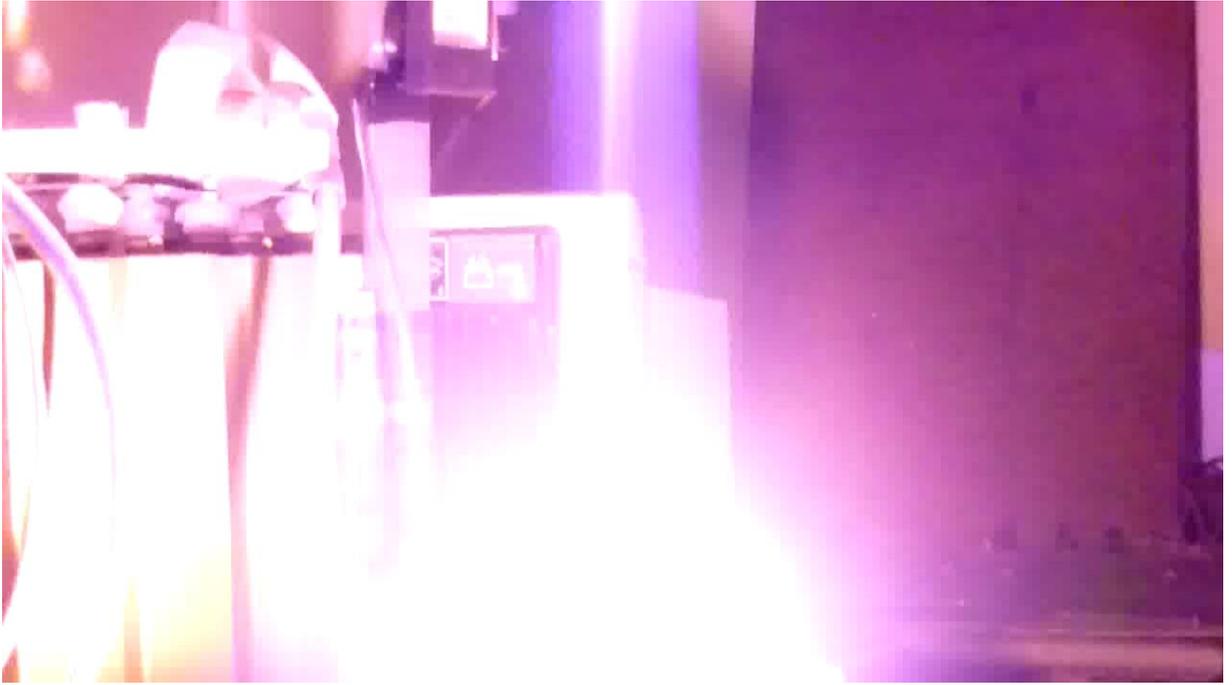
Fig.5

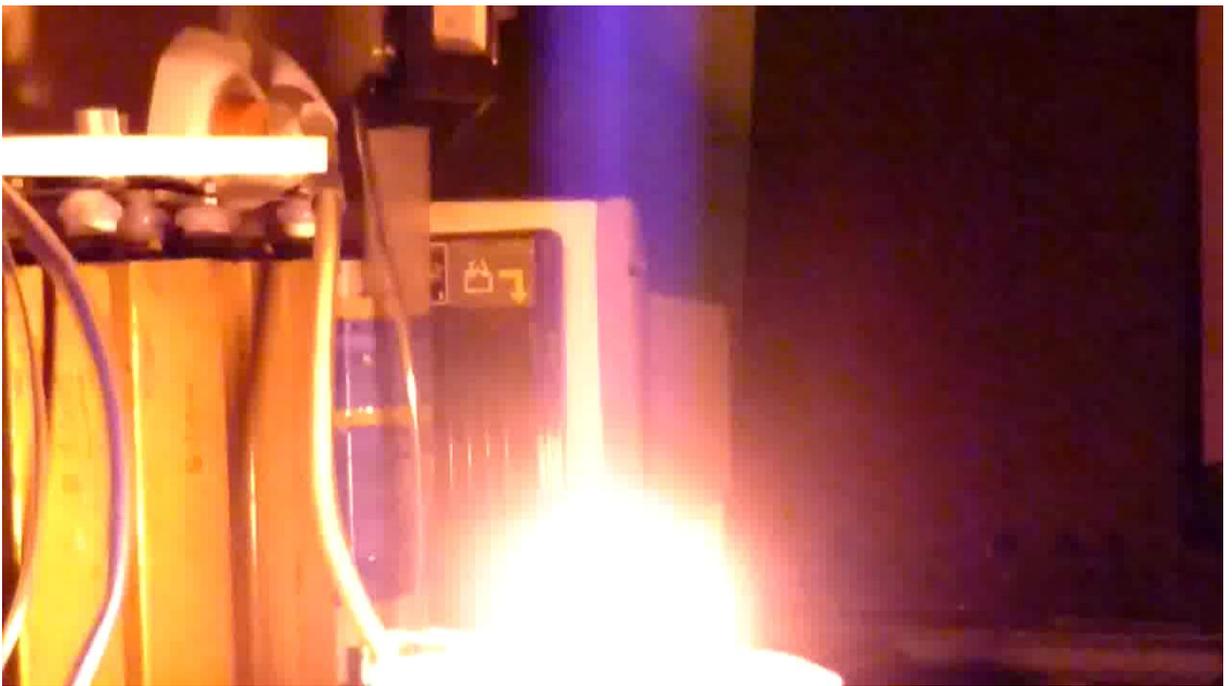
Fig.6

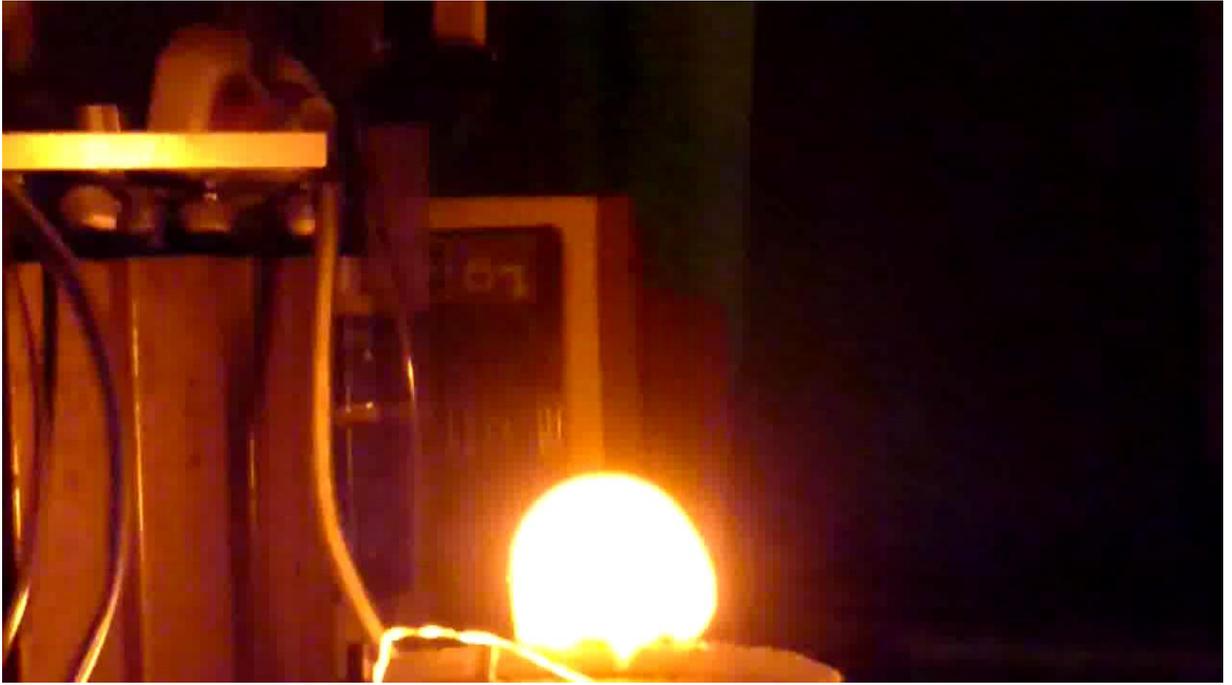

Fig.7

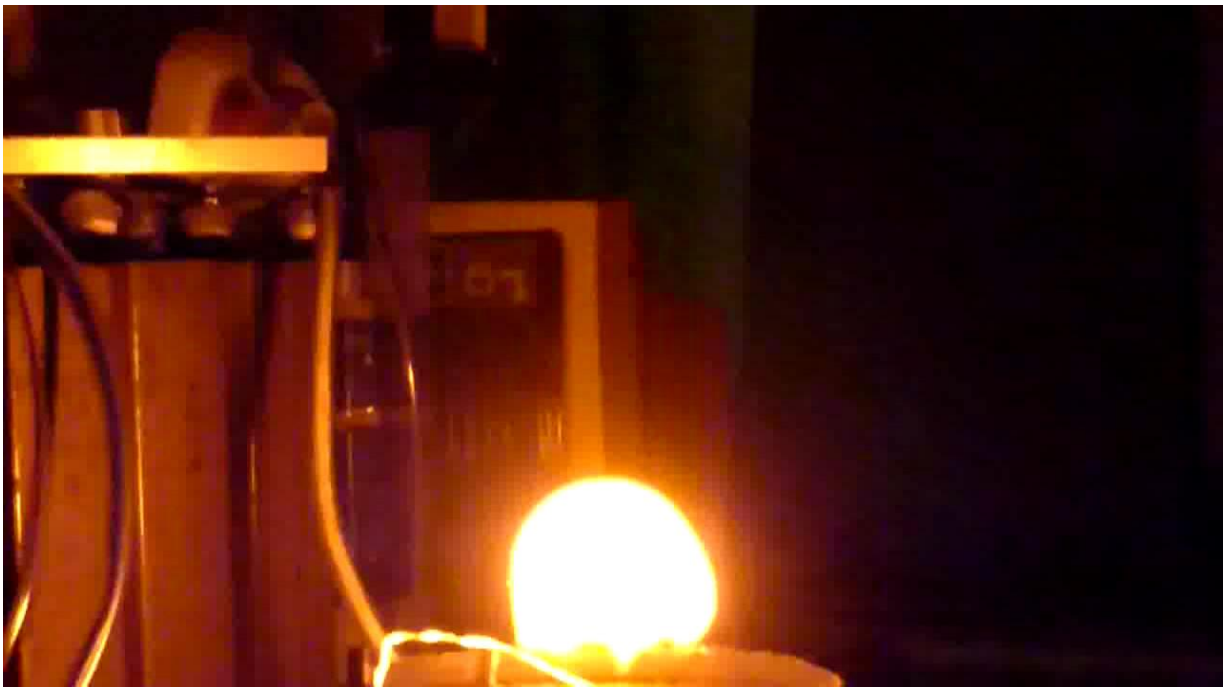

Fig.8

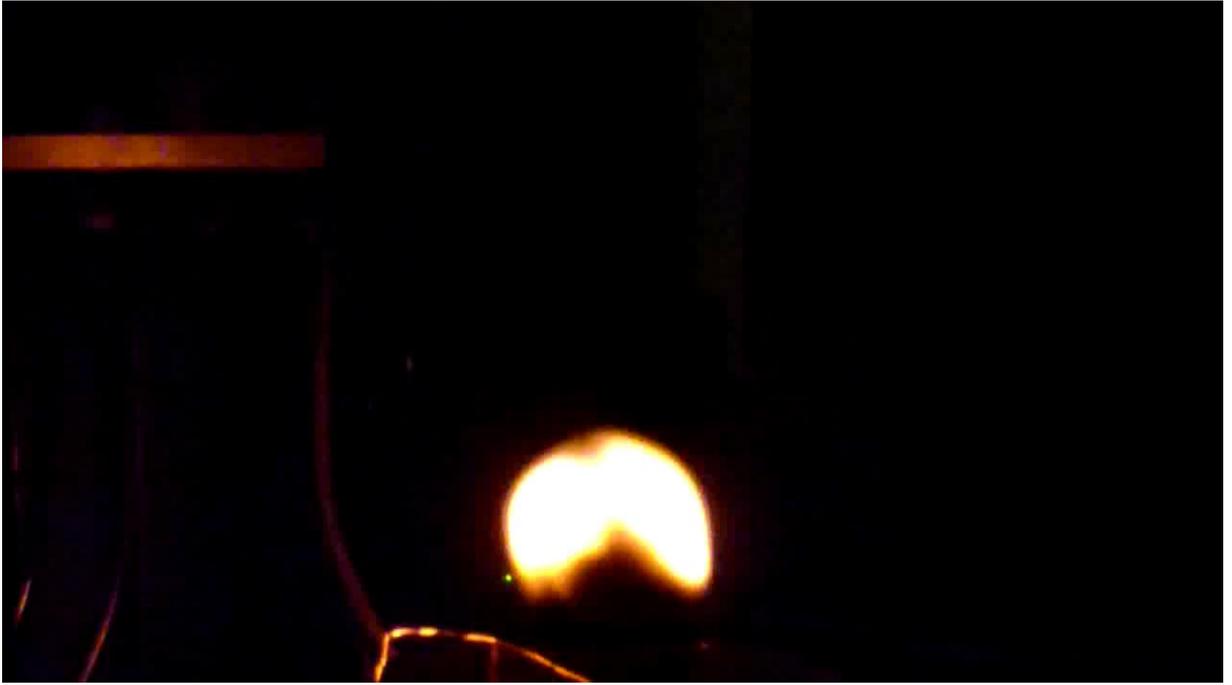

Fig.9

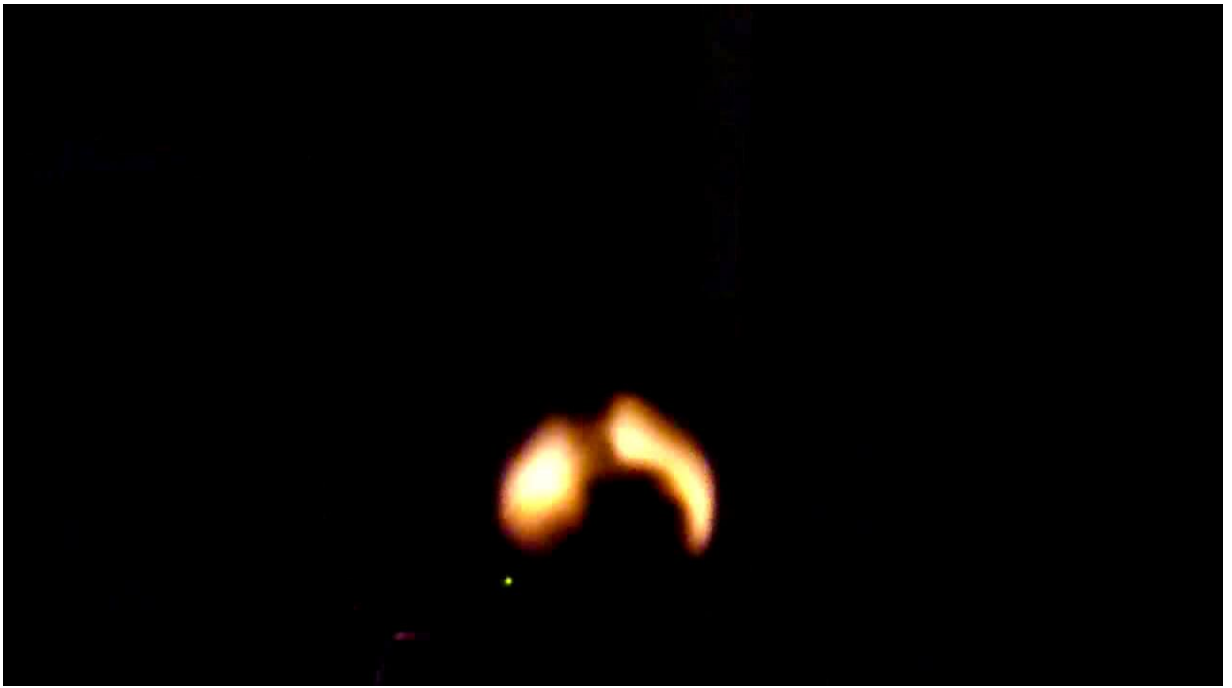

Fig.10

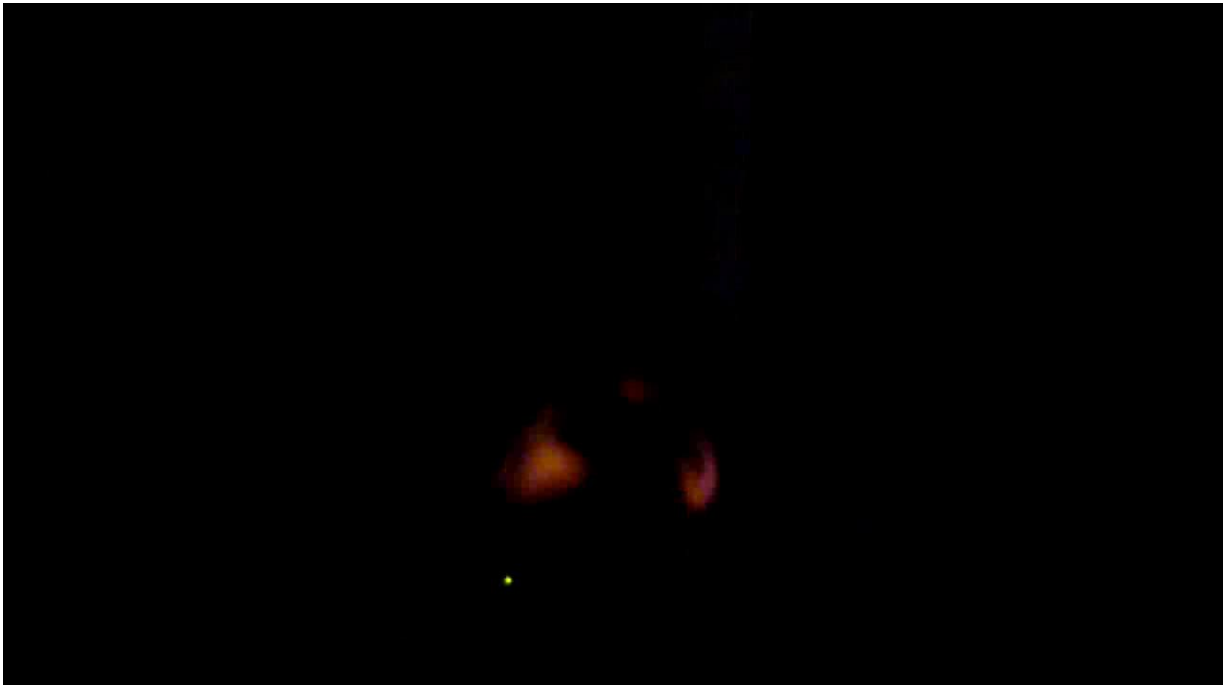

Fig. 11. Figures: 1-11: some elements of the design of the experimental setup and the processes of generation and decay of the ball-shaped luminous cloud self-assembled of the unipolar charged, short-circuited thermionic batteries, i.e., the unipolar charged aerosol particles/droplets, continuously emitting thermoelectrons and continuously re-absorbing the thermoelectrons and negative air ions, are shown in a series of the frames. The relatively long-living luminous decomposing clouds ($\tau \leq 1s$) were generated by us with help of the high-voltage arc discharge (U ~ 8 - 10kV, I ~ 100 - 150 A) initiated on the surface of a wide variety of either the individual liquids or high reactive compositions prepared in situ by the synchronous arc evaporation of the several pre-separated liquid components, in particular, on the surface of the highly concentrated, highly reactive water solutions of hydrogen peroxide, the solutions of sulfuric acid, nitric acid, hydrochloric acid, formic acid, performic acid (in situ syntesized by the co-evaporation and reactive co-condensation of the vapor of formic acid and hydrogen peroxide), acetic acid, the solutions of methyl nitrate (in situ syntesized by the co-evaporation and reactive co-condensation of the vapor of methanol and nitric acid), the solutions of ammonium nitrate, phenols, etc.